\documentclass[fleqn]{article}
\usepackage[dvips]{graphicx}

\author{L.Chotorlishvili, Z.Toklikishvili, V.Bochorishvili , A.Sagaradze}
\date{}
\title{\large \bf Investigation of Quantum Chaos in the Parametric Dependent System of
Interacting oscillators }

\begin{document}
\maketitle
\begin{center}
Tbilisi State University, Georgia, 0128. Tbilisi, Chavchavadze av.
3 \\
Email: lchotor@yahoo.com
\begin{abstract}
Formation of chaos in the parametric dependent system of
interacting oscillators for the both classical and quantum cases
has been investigated. Domain in which classical motion is chaotic
is defined. It has been shown that for certain values of the
parameters from this domain, form of the classical power spectrum
is in a good agreement with the quantum band profile. Local
density of states is calculated. The range in which application of
perturbation theory is correct has been defined.
\end{abstract}
Pacs number 05.45. -a
\end{center}
\section{ Introduction. Problem statement.}
At last time the investigation of parametric dependent
Hamiltonians like $H(Q,P,\lambda)$ is a subject of great interest.
Here $(Q,P)$ is a set of canonical coordinates and impulses, and
$\lambda$ is a parameter describing connection of the system with
the external field. Interest to such systems is motivated by
mesoscopic physics [1]. Most of the works devoted to the
parametric dependent systems are dealing with the investigation of
following situation: For $\lambda=0$ Hamiltonian is exactly
integrable. For $\lambda>0$ Hamiltonian $H(Q,P,\lambda)$ becomes
non-integrable and for some defined values of
$\lambda=\lambda_{0}$, solutions of classical equations generated
by the Hamiltonian $H(Q,P,\lambda_{0})$ display chaos. After this,
the little variation of the parameter
$\delta\lambda=\lambda-\lambda_{0}$ is of interest. Using the
basis $\left|\Phi_{n}\right\rangle$ in which
$\hat{H}(Q,P,\lambda_{0})$ is diagonal, for small values of
$\delta\lambda$, Hamiltonian $\hat{H}(Q,P,\lambda)$ can be written
in the form of sum of two matrix $$ \hat{H}=E_{0}+\delta\lambda
B.$$ Here $E_{0}$ is the diagonal matrix, and $B$ is the banded
matrix, off-diagonal elements of which are random numbers [2]. By
way of numerical diagonalization it is possible to define
eigenvectors
$\left|\Phi_{n}(\lambda_{0}+\delta\lambda)\right\rangle$,\,
$\left|\Phi_{n}(\lambda_{0})\right\rangle$ and eigenvalues
$E_{n}(\lambda_{0}+\delta\lambda)$,\,$E(\lambda_{0})$
corresponding to the Hamiltonians
$\hat{H}(Q,P,\lambda)$,\,$\hat{H_{0}}(Q,P,\lambda_{0})$
appropriately. The main interest at this is connected with the
dependence of parametric kernel
$$P(n|m)=\left|\left\langle\Phi_{n}(\lambda_{0}+\delta\lambda)|\Phi_{m}(\lambda_{0})\right\rangle\right|^{2}$$
from the parameter $\delta\lambda$.  The value of
$P(r)=\overline{P(n|n+r)}$ averaged over the $n$ can be considered
as the local density of states. Descriptions of foregoing methods
of solving of problems of quantum chaos are given in [3-5]. Goal
of this paper is investigation of chaos in the system of
interacting oscillators when both of the Hamiltonians
$\hat{H_{0}}(Q,P,\lambda_{0})$, $\hat{H}(Q,P,\lambda)$ display
chaos. Situation when the Hamiltonian $\hat{H_{0}}$ is integrable
has been studied by us earlier (see papers [6,7]).
\newpage
\section{ Classical consideration. Model Hamiltonian.}
Investigation of the following Hamiltonian is of interest for us
 \begin{eqnarray}   \label{1}
   H&=&H_{0}+\lambda V \nonumber ,\\
  H_{0}&=&\frac{P_{1}^{2}}{2}+\frac{P_{2}^{2}}{2}+\frac{P_{3}^{2}}{2}+
 \frac{1}{2}(q_{1}^{2}+q_{2}^{2}+q_{3}^{2}) ,\\
  V&=&q_{1}^{2}q_{2}^{2}q_{3}^{2} \nonumber.
 \end{eqnarray}

Here and below we  work with the dimensionless quantities.
Canonical equations for the Hamiltonian (\ref{1}) are of form
\begin{eqnarray}\label{2}
\dot{q_{1}}&=&p_{1} \nonumber ,\\
\dot{p_{1}}&=&-q_{1}-2\lambda q_{1}q_{2}^{2}q_{3}^{2} ,\nonumber \\
\dot{q_{2}}&=&p_{2}  ,\\
\dot{p_{2}}&=&-q_{2}-2\lambda q_{2}q_{1}^{2}q_{3}^{2} ,\nonumber \\
\dot{q_{3}}&=&p_{3} \nonumber ,\\
\dot{p_{3}}&=&-q_{3}-2\lambda q_{3}q_{1}^{2}q_{2}^{2} \nonumber.
\end{eqnarray}
Solutions for the set of equations in case of  $\lambda\neq0$ can
be obtained only numerically. As is shown by numerical analyze,
solutions of the set of equations  (\ref{2}) for the values of the
parameters $\lambda=3.0$, $H=4.2$ are essentially chaotic (see
Fig.1). For further calculations we have to obtain classical power
spectrum with the purpose of its further comparison with the band
profile. Let us construct correlation function for
$V(q_{1}\,q_{2}\,q_{3})=-\frac{\partial H}{\partial\lambda}$
\begin{equation}\label{3}
G(\tau)=\left\langle\left(V(t+\tau)-\bar{V}\right)\left(V(t)-\bar{V}\right)\right\rangle
\end{equation}
where $\left\langle\ldots\right\rangle$ means averaging over the
time, $\bar{V}$ is the averaged values of $V$. We have to make
Fourier transformation of the expression (\ref{3}) to get
classical power spectrum
\begin{equation}\label{4}
C(\omega)=\int e^{-i\omega t}G(\tau)d\tau.
\end{equation}
In the practical calculation of (\ref{4}) we use the method
equivalent of (\ref{4})
\begin{equation}\label{5}
C(f)=\frac{1}{T}\left|\int_{-T/2}^{T/2}e^{-i\omega t}V(t)dt\right|,
\end{equation}
here $x \in \left[0,T \right]$ is time interval in which solution
of the set of equations (\ref{2}) is obtained,
$f=\frac{\omega}{2\pi}$. To get the power spectrum one needs to
use discrete form of expression (\ref{5})
\begin{equation} \label{6}
C(k)=\frac{T\Delta}{2l+1}\sum_{n=-l}^{l}\left|X(k+n)\right|^{2},
\end{equation}
here $X(k)=\sum_{j=0}^{N-1}V(j)e^\frac{i 2 \pi k}{N\Delta}$  ,
$k=fN\Delta$, $N$ is a number of dots dividing the interval $T$,
$\Delta=\frac{T}{N}$, $l$ is the averaging parameter [8,9]. Form
of the classical power spectrum
 for the different values of the parameter $l$ is given on the Figures Fig. 2, Fig.3.
 These data are confirming the fact that solutions of (\ref{2}) are chaotic.
\newpage
 \section{Quantum mechanical consideration}
 Our subsequent goal is the comparison of the band profile $\left|B_{nm}\right|^{2}$ considered as the
 function of $\omega=\frac{E_{n}-E_{m}}{\hbar}$ with the classical band profile. Let us set about to the quantum-mechanical
 analyze of the Hamiltonian (\ref{1}). Eigenvectors $ \left|\Psi_{n}\right\rangle$ of the basis, in which Hamiltonian
 $H(P,Q,0)$ is diagonal can be found easily and are of the form [10]
 \begin{equation}\label{7}
 \Psi_{n_{1}n_{2}n_{3}}(q_{1}q_{2}q_{3})=\frac{H_{n_{1}}(q_{1})H_{n_{2}}(q_{2})H_{n_{3}}(q_{3})}{\sqrt{2^{n_{1}+n_{2}+n_{3}}n_{1}!n_{2}!n_{3}!\pi^{3}}}e^{-\frac{1}{2}(q_{1}^{2}+q_{2}^{2}+q_{3}^{2})},
 \end{equation}
where $H_{n}(q)$ denotes the Hermite polynomials [11]. Eigenvalue
energy is given by the formula
\begin{equation}\label{8}
E_{n}=(n+\frac{3}{2}),
\end{equation}
where
\begin{equation}\label{9}
n=n_{1}+n_{2}+n_{3}.
\end{equation}
Condition (\ref{9}) implies existence of the
$\frac{(n+1)(n+2)}{2}$ fold degeneration for the given $n$.
\newline Diagonalization of the Hamiltonian (1) can be done only
numerically. On the Figures 4, and 5 dependence of the energy
eigenvalues $E_{n}(\lambda)$ from the parameter $\lambda$ are
plotted. According to Fig.4 for the $\lambda\neq0$ removal of
degeneration takes place. In case of further increasing of the
parameter $\lambda$ up to the values characterizing the domain of
chaotic motion  $\lambda=3.0$, as easy to see from Fig.5,
repulsion of the energy terms happens. This fact is of great
importance since the repulsion of the energy terms is a sine of
emerging of quantum chaos [12].
 For the defining of the band profile at first we need to find eigenvectors of the
 Hamiltonian $H(P,Q,\lambda_{0})$. This problem is possible to resolve by way of numerical diagonalization
 of the Hamiltonian $H(P,Q,\lambda_{0})$, at the values of
 $\lambda_{0}=3.0$ in the basis of functions (\ref{7})
\begin{equation}\label{10}
\left\langle\Psi_{n_{1}n_{2}n_{3}}\left|\left(H_{0}+\lambda_{0}V\right)\right|\Psi_{m_{1}m_{2}m_{3}}\right\rangle
\end{equation}
Since the matrix elements are satisfying the condition
$$\left\langle n\left|V(q_{1}q_{2}q_{3})\right|m\right\rangle=\left\langle n_{1}n_{2}n_{2}\left|V(q_{1}q_{2}q_{3})\right|m_{1}m_{2}m_{2}\right\rangle=\left\langle n_{1}\left|q_{1}^{2}\right|m_{1}\right\rangle\left\langle n_{2}\left|q_{2}^{2}\right|m_{2}\right\rangle\left\langle
n_{3}\left|q_{3}^{2}\right|m_{3}\right\rangle,$$
 taking into account that
 \begin{equation}\label{11}
 q_{nm}=\frac{1}{\sqrt{2}}(\sqrt{n}\delta_{n-1,m}+\sqrt{n+1}\delta_{n+1,m}),
\end{equation}
and using the condition
$\left(q^{2}\right)_{nm}=\sum_{k}q_{nk}q_{km}$ we get
\begin{equation}\label{12}
\left(q^{2}\right)_{nm}=\frac{1}{2}(\sqrt{n}\sqrt{n-1}\delta_{n-2,m}+(2n+1)\delta_{n,m}+\sqrt{n+1}\sqrt{n+2}\delta_{n+2,m}).
\end{equation}
In (\ref{11}) and (\ref{12})  $\delta_{n,m}$ stands for Kronecker
delta function.
 Substituting (\ref{11}) and (\ref{12}) into (\ref{10}) after
straightforward calculations one can obtain
\begin{eqnarray}
\left\langle
n_{1}n_{2}n_{3}\left|V\right|m_{1}m_{2}m_{3}\right\rangle&=&\frac{1}{8}(\sqrt{n_{1}}\sqrt{n_{1}-1}\delta_{n_{1}-2,m_{1}}+(2n_{1}+1)\delta_{n_{1},m_{1}}+\sqrt{n_{1}+1}\sqrt{n_{1}+2}\delta_{n_{1}+2,m_{1}})\times \nonumber \\
                                                         & & \mbox{} \times(\sqrt{n_{2}}\sqrt{n_{2}-1}\delta_{n_{2}-2,m_{2}}+(2n_{2}+1)\delta_{n_{2},m_{2}}+\sqrt{n_{2}+1}\sqrt{n_{2}+2}\delta_{n_{2}+2,m_{2}})\times \nonumber \\
                                                         & & \mbox{} \times(\sqrt{n_{3}}\sqrt{n_{3}-1}\delta_{n_{3}-2,m_{3}}+(2n_{3}+1)\delta_{n_{3},m_{3}}+\sqrt{n_{3}+1}\sqrt{n_{3}+2}\delta_{n_{3}+2,m_{3}}) ,\nonumber
\end{eqnarray}
\begin{equation}\label{13}
\left\langle
n_{1}n_{2}n_{3}\left|H_{0}\right|m_{1}m_{2}m_{3}\right\rangle=(n_{1}+n_{2}+n_{3}+\frac{3}{2})\delta_{n_{1}m_{1}}\delta_{n_{2}m_{2}}\delta_{n_{3}m_{3}}.
\end{equation}
After forming from the matrix elements (\ref{13}) matrix and
making diagonalization one can obtain eigenvalues spectrum
$E_{n(\lambda_{0})}$ and basis of eigenvectors
$\Phi_{n}(\lambda_{0})$. Calculating matrix elements of
$V(q_{1},q_{2},q_{3})$ in the basis $\Phi_{n}(\lambda_{0})$ we can
get the band profile
\begin{equation} \label{14}
B_{nm}^{2}=\frac{4\pi^{2}\hbar^{2}}{\triangle_{0}}\left|\left\langle\Phi_{n}(\lambda_{0})\left|V(q_{1},q_{2},q_{3})\right|\Phi_{m}(\lambda_{0})\right\rangle\right|^{2}
\end{equation}
Here $\Delta_{0}=a \hbar^{d}$ is the mean level spacing, and $d$
is the system's dimension. Coefficient $a$ is determined from the
condition  $a=\frac{1}{D(E)}$ , where $D(E)=\frac{\delta N}{\delta
E}$ is the density of states, and $\delta N$ is the number of
states in the energy interval  $\delta E=E(P,Q,\lambda_{0}+\delta
\lambda)-E(P,Q,\lambda_{0})$. Density of states in it's turn is
possible to determine from the numerical diagonalization of the
Hamiltonian $H(P,Q,X_{0})$ , or from the integral
\begin{equation}
D(E)=\frac{1}{(2\pi\hbar)^{\alpha}}\frac{\partial}{\partial
E}\int_{H(P,Q,\lambda)=E}dPdQ .
\end{equation}
Here the integral must be taken over the phase space enclosed by
energy surface (see Fig.6) In our calculations we are interested
in the domain in which classical motion is essentially chaotic:
$E(P,Q,\lambda=3)=4.2;$ $\delta\lambda=0.1;$ $\delta E=0.1;$
$\hbar=0.1;$ $a\approx7.2$. Taking smaller values of $\hbar$ leads
to the necessity of diagonalization of matrix larger then our. In
our case ($\hbar=0.1$) matrix dimension is $455\times455$. Band
profile $B_{nm}^{2}$ as a function of
$\omega=\frac{E_{n}-E_{m}}{\hbar}$ is plotted on the Fig.7.
According to Fig.7 band profile is in a good qualitative agreement
with the classical power spectrum. Better quantitative agreement
is possible to achieve for smaller values of $\hbar$ [1] (since
the quasi-classical approximation implies $\hbar\rightarrow 0$).
Another quantity we are interested in is the parametric kernel
\begin{equation}\label{16}
P(r)=\overline{\left|\left\langle\Phi_{n+r}(\lambda_{0}+\delta\lambda)|\Phi_{n}(\lambda_{0})\right\rangle\right|^{2}},
\end{equation}
where the averaging is done over the several $n$ states. Evidently
(\ref{16}) determines correlation between the wave functions. The
parametric kernel (\ref{16}) frequently is named as the Local
Density of States (LDS) [3]. We are interested in $P(r)$, since
with its help, is possible to determine the boundaries in which
perturbation theory is valid. For that one has to compare
parametric kernel (\ref{16}) with the results for LDS obtained by
means of perturbation theory [5]
\begin{equation} \label{17}
P_{prt}(r)=\frac{\delta
x^{2}\left|B_{nm}\right|^{2}}{\Gamma^{2}(\delta
x)+(E_{n}-E_{m})^{2}},
\end{equation}
where the parameter $\Gamma(\delta x)$ for given $\delta x$  must
be defined from the normalization condition
\begin{equation} \label{18}
\sum_{r}P_{prt}(r)=1 .
\end{equation}
Results of numerical calculations for the values $\delta x=0.1$
and $\delta x=0.2$ are presented on the Figures Fig.8 and Fig.9.
On basis of obtained results we conclude that boundary in which
perturbation theory for our system is correct is limited by the
condition  $\delta x\leq1$.

\newpage

\begin{figure}
\begin{center}
\includegraphics[angle=0, width=0.5\textwidth  ]{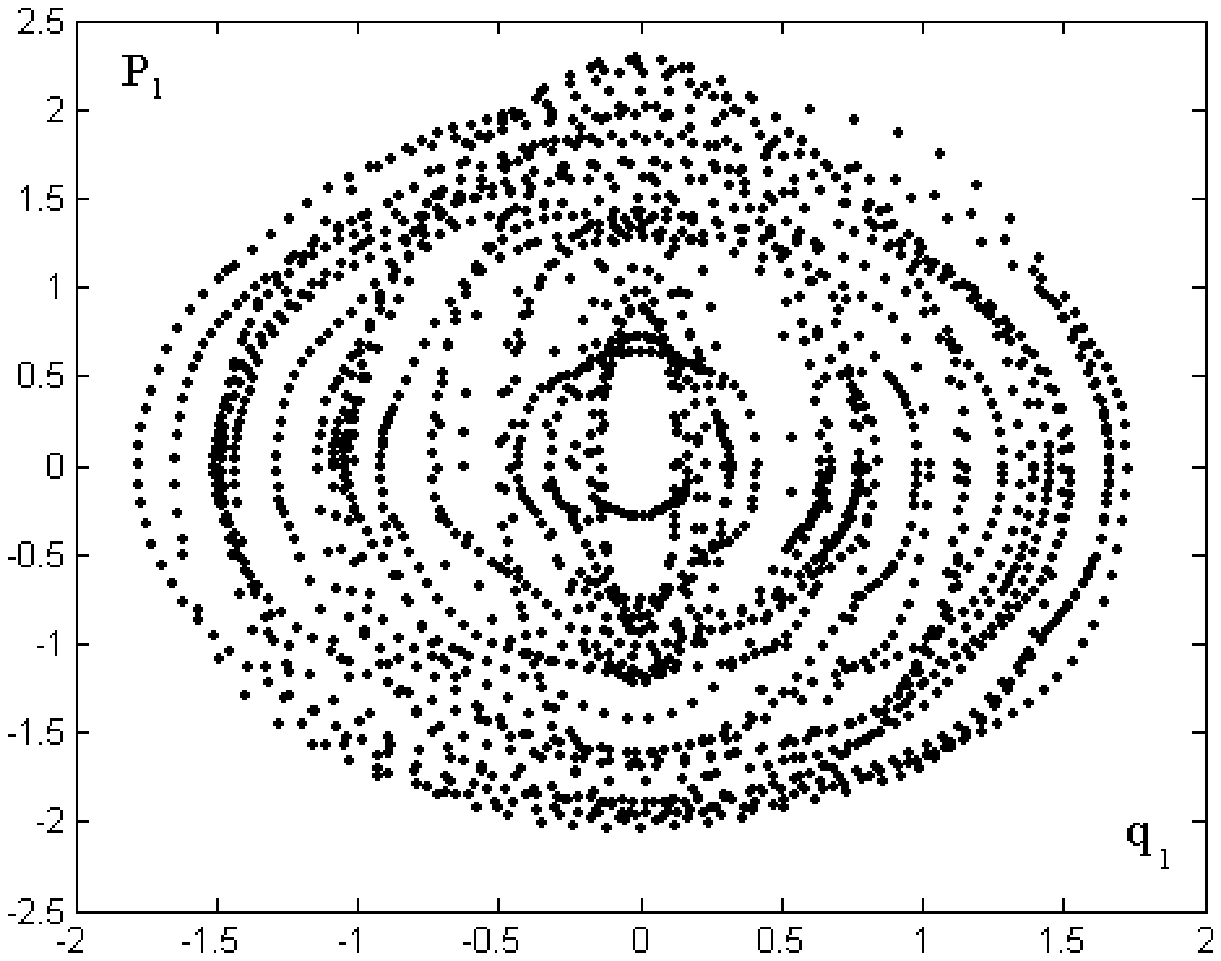}
\newline Fig.1 Phase trajectory projection on the plane $(p_{1},
q_{1})$ obtained for the values of the   parameters: $E=4.2,
\lambda=3.0$. It is easy to see that trajectory is chaotic.
\end{center}
\end{figure}

\begin{figure}
\begin{center}
\includegraphics[angle=0, width=0.5\textwidth  ]{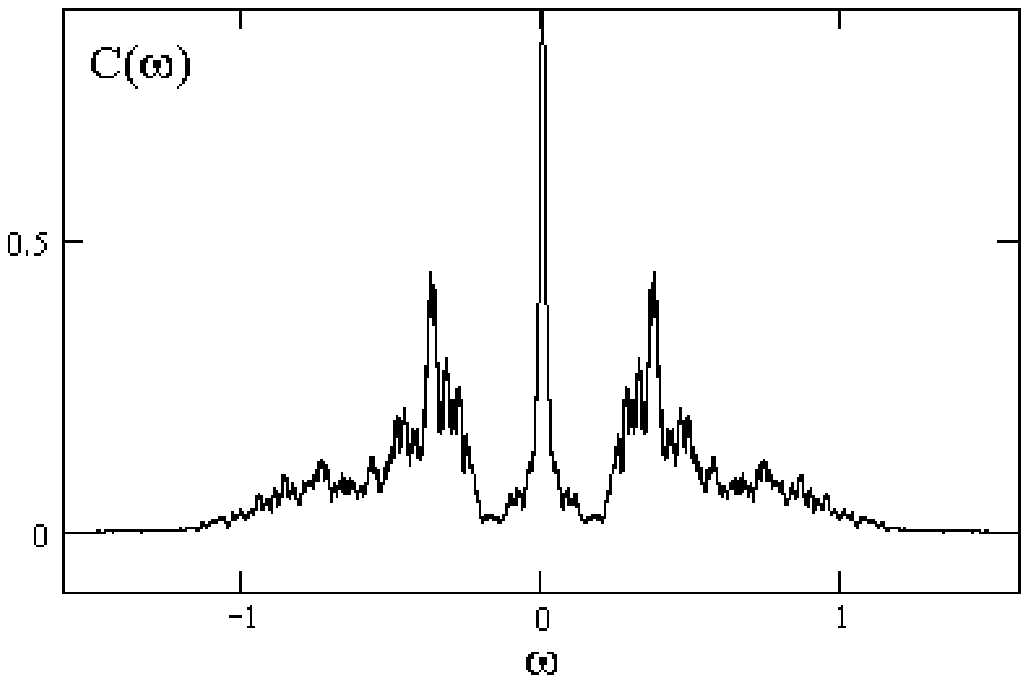}
\newline Fig.2  Classical Power Spectrum plotted for the values of the
parameters: $E=4.2, \lambda=3.0 ,l=10$
\end{center}
\end{figure}

\newpage

\begin{figure}
\begin{center}
\includegraphics[angle=0, width=0.5\textwidth  ]{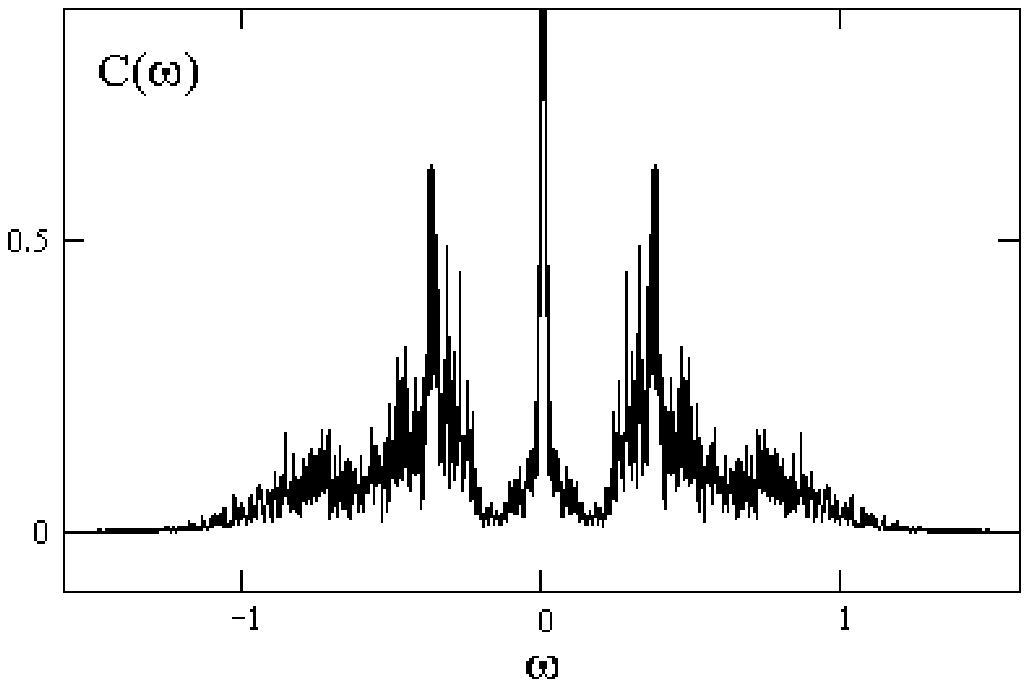}
\newline Fig.3  Classical Power Spectrum plotted for the values of the
parameters: $E=4.2, \lambda=3.0, l=3$
\end{center}
\end{figure}

\begin{figure}
\begin{center}
\includegraphics[angle=0, width=0.5\textwidth  ]{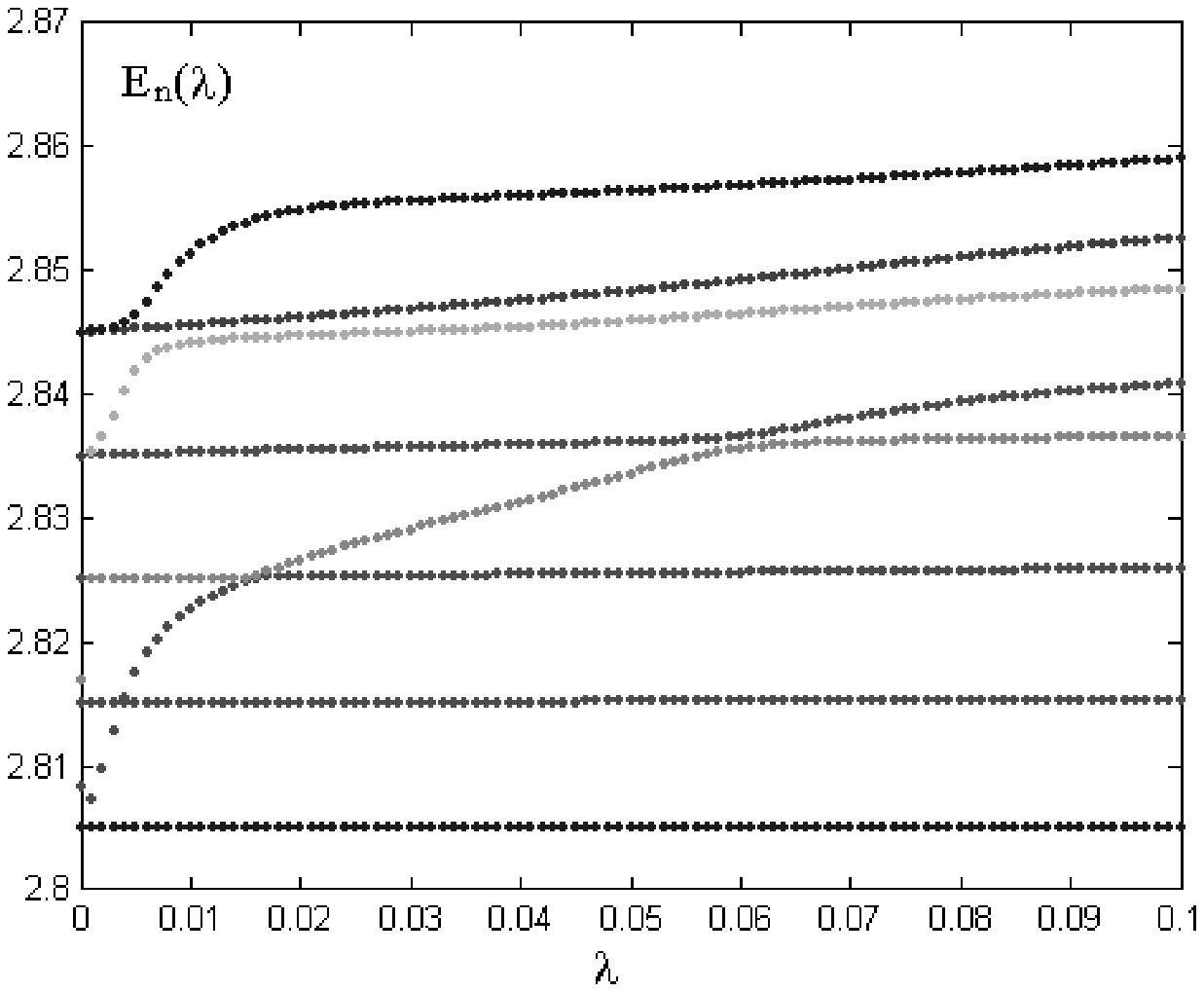}
\newline Fig. 4 Energy terms as a functions of the parameter $\lambda$
\end{center}
\end{figure}

\newpage

\begin{figure}
\begin{center}
\includegraphics[angle=0, width=0.5\textwidth  ]{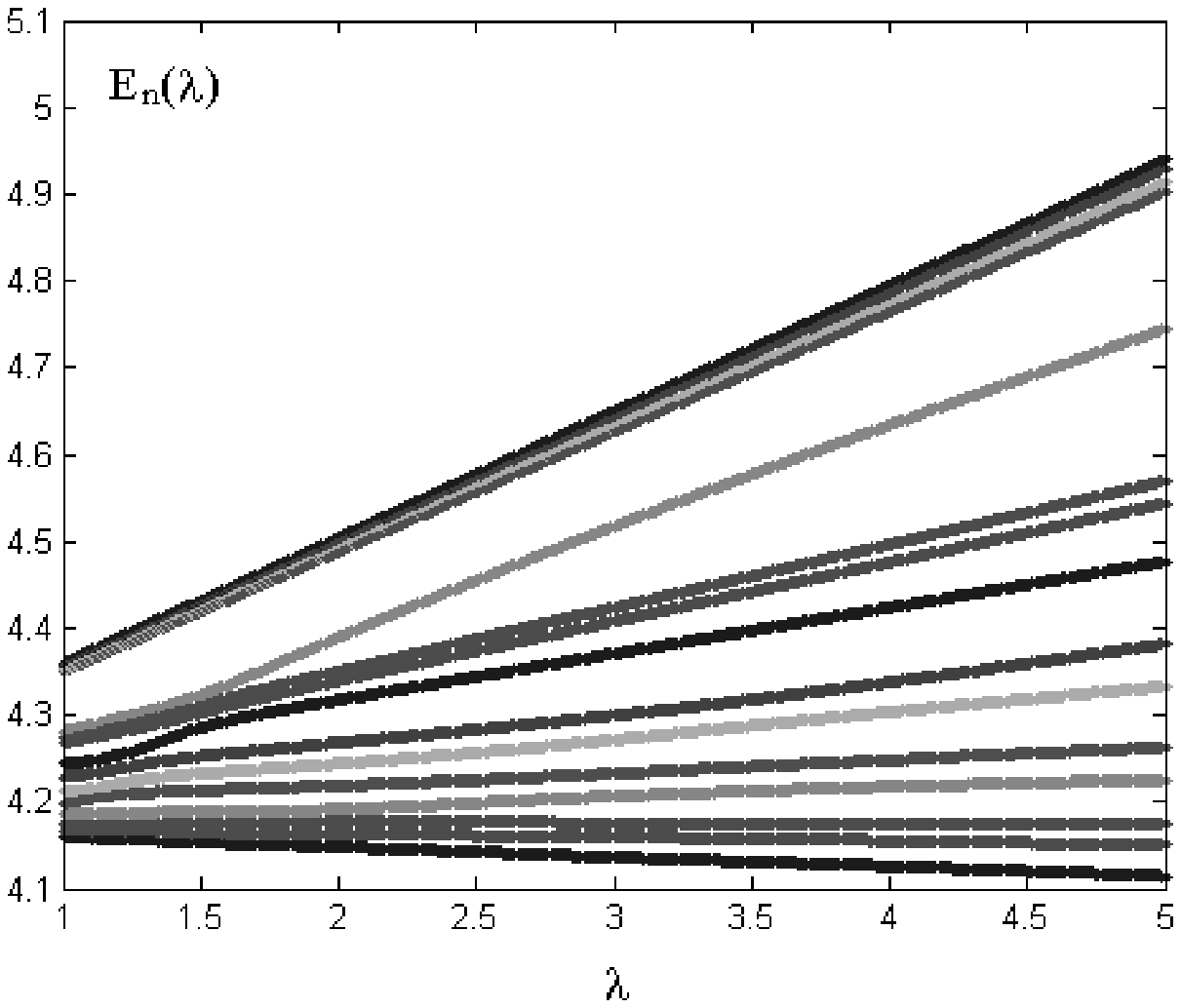}
\newline Fig. 5 Energy terms as a functions of the parameter $\lambda$. For
the values from the chaotic domain $\lambda\geq3.0$, repulsion of
the energy levels happens.
\end{center}
\end{figure}

\begin{figure}
\begin{center}
\includegraphics[angle=0, width=0.5\textwidth  ]{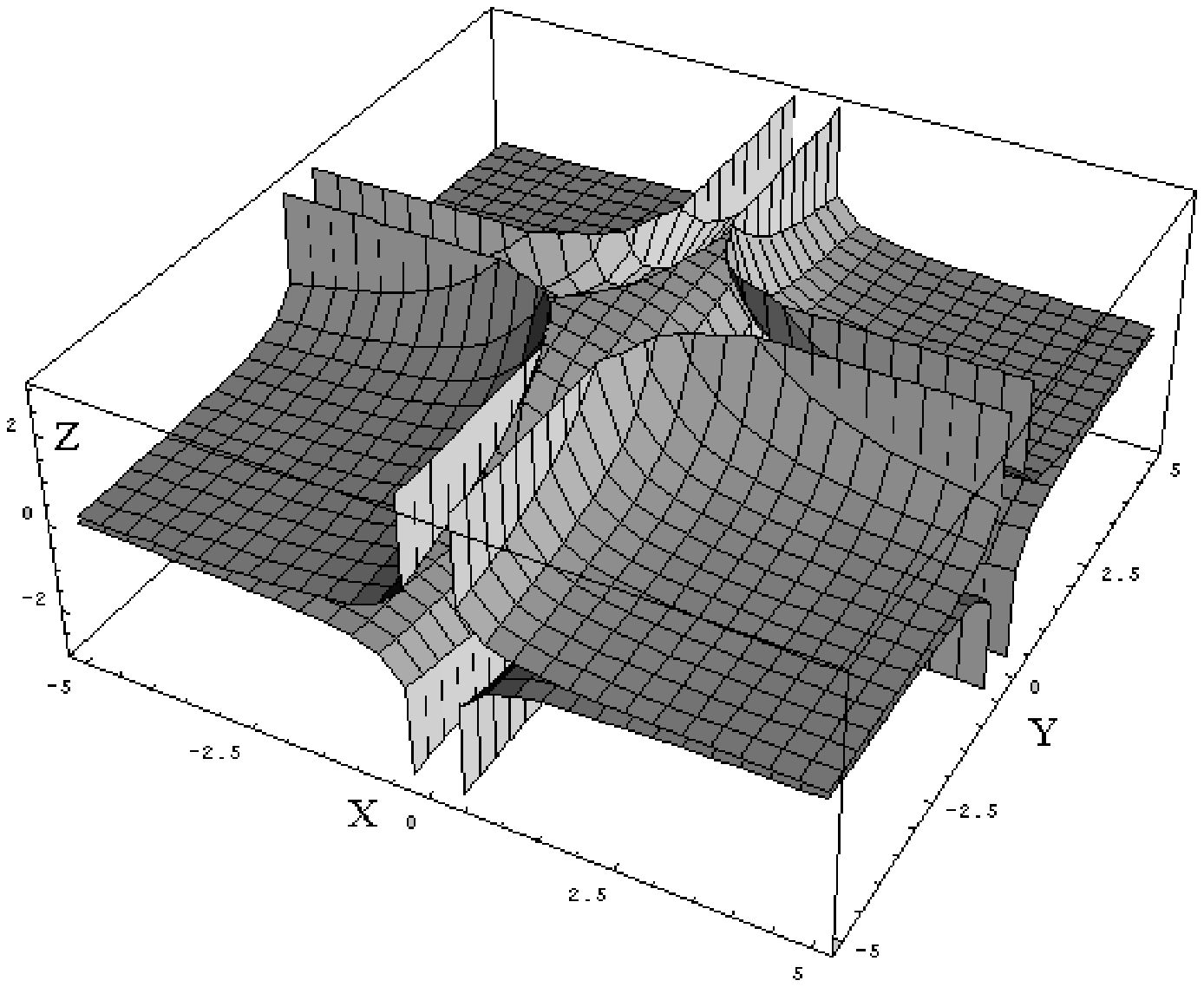}
\newline Fig. 6 Equipotential surface constraint by the condition
$H(P,Q,\lambda)=E=4,2$
\end{center}
\end{figure}

\newpage

\begin{figure}
\begin{center}
\includegraphics[angle=0, width=0.5\textwidth  ]{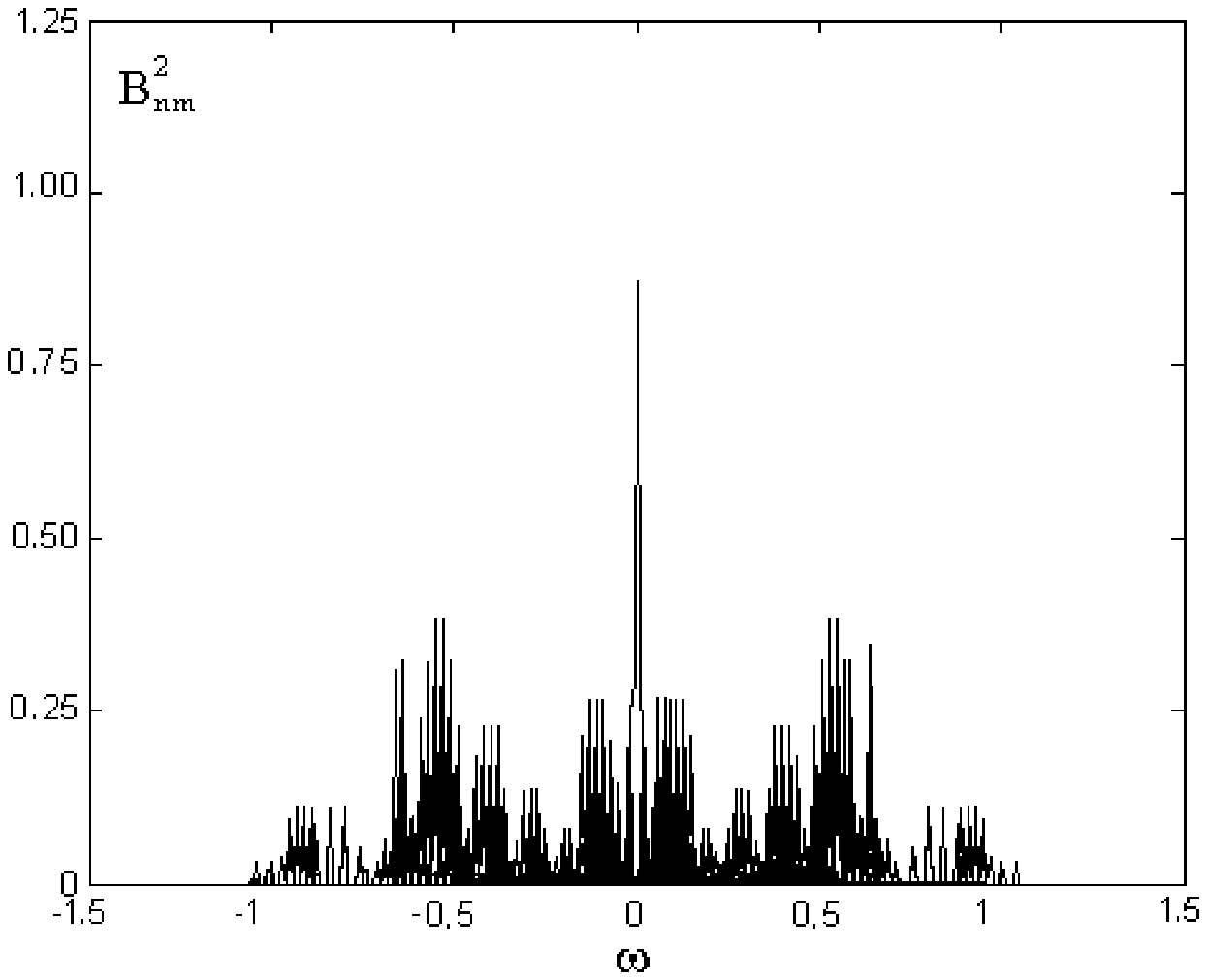}
\newline Fig. 7 Quantum band profile  $B_{nm}^{2}$ as a function of
$\omega=\frac{E_{n}-E_{m}}{\hbar}$ for the values of parameters
$E=4.2, \lambda=3.0 ,\delta\lambda=0,1$. Evidently $B_{nm}^{2}$ is
in a good qualitative agreement with the Classical Power Spectrum
(see Fig. 2 )
\end{center}
\end{figure}

\begin{figure}
\begin{center}
\includegraphics[angle=0, width=0.5\textwidth  ]{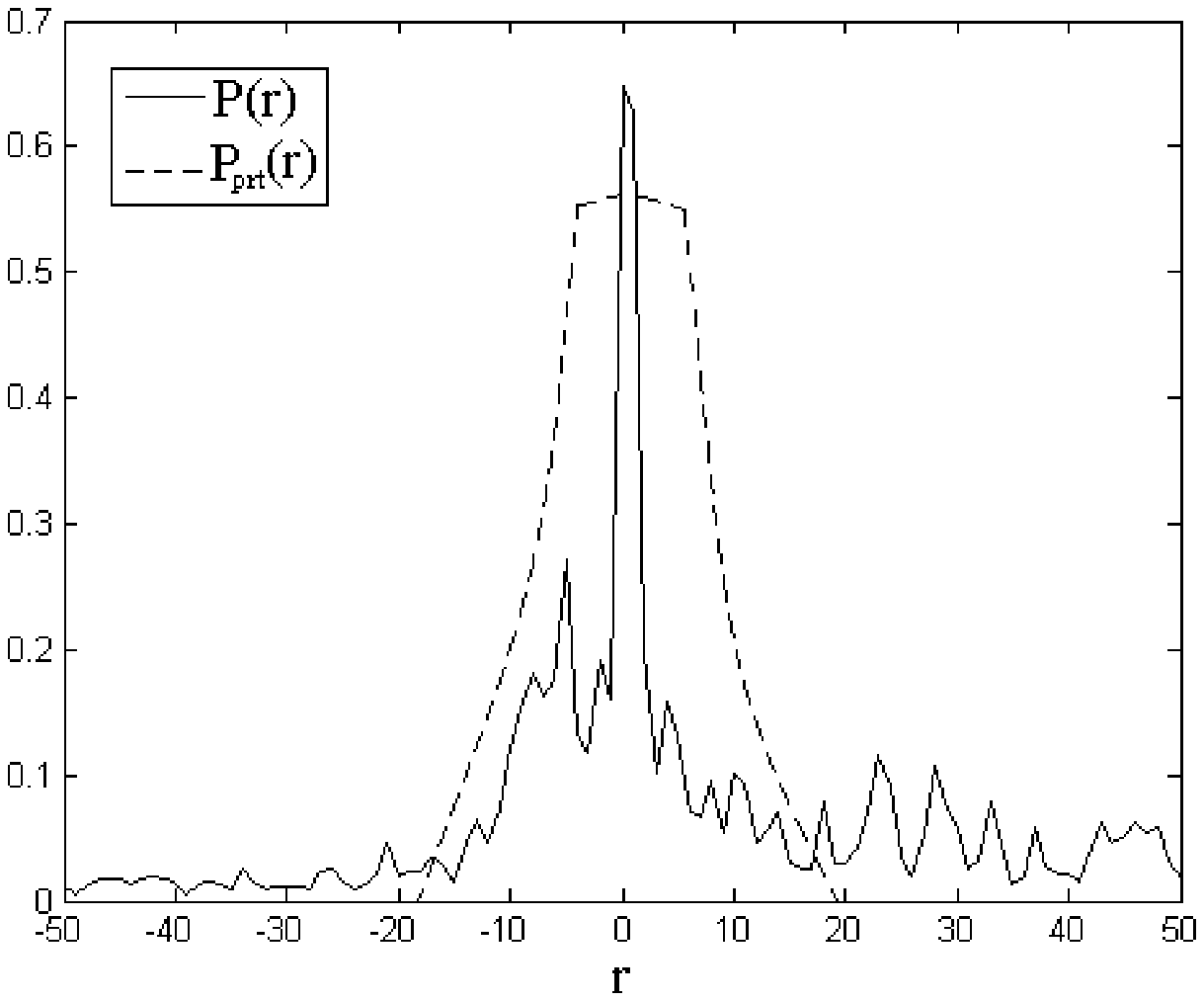}
\newline Fig. 8 Quantum Profile $P(r)$ (Local Density of States) and
First-Order Perturbative Profile $P_{prt}(r)$  for the values of
the parameter $\lambda=0,1$.
\end{center}
\end{figure}

\newpage

\begin{figure}
\begin{center}
\includegraphics[angle=0, width=0.5\textwidth  ]{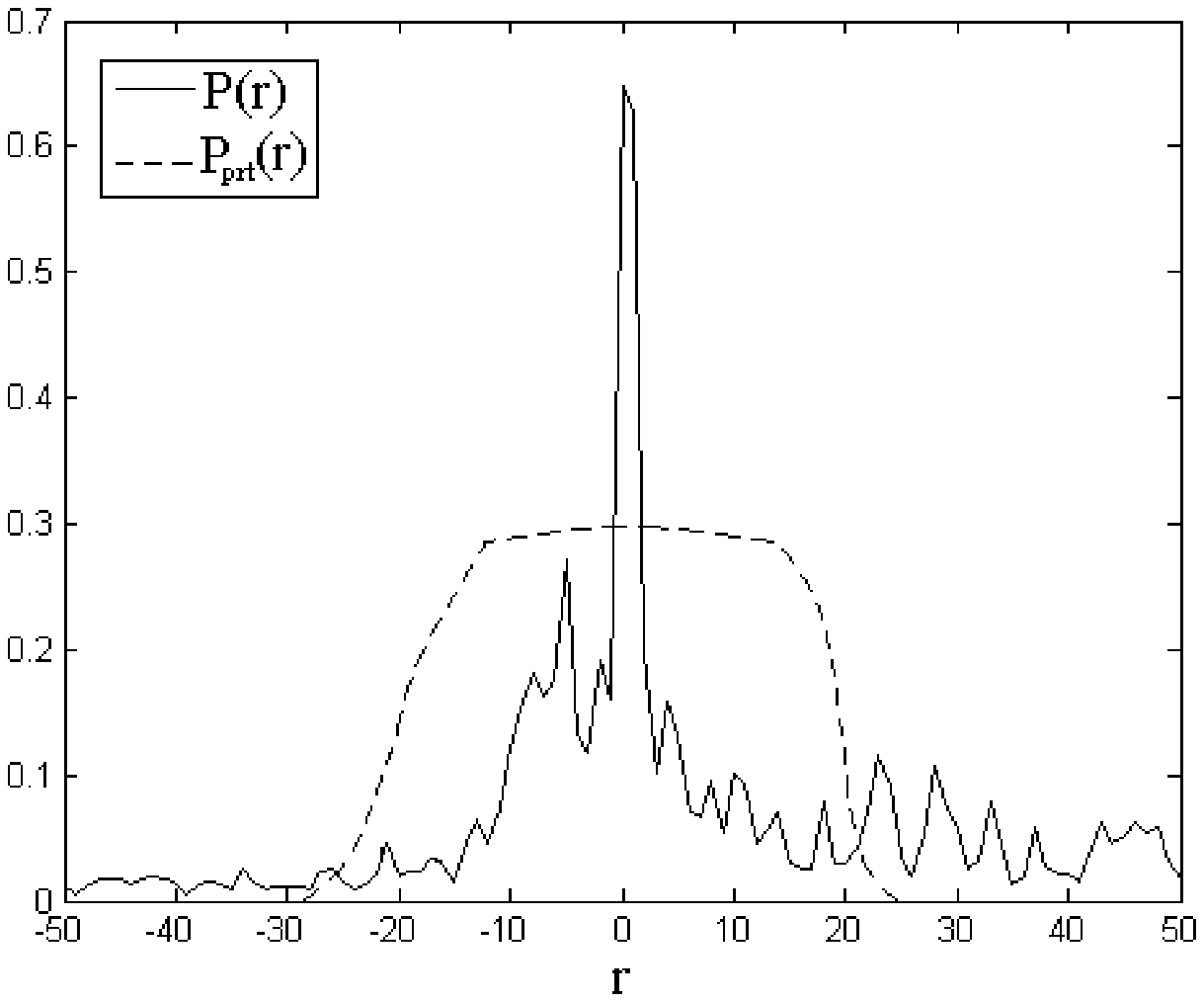}
\newline Fig. 9 Quantum Profile $P(r)$ (Local Density of States) and
First-Order Perturbative Profile $P_{prt}(r)$  for the values of
$\lambda=0,2$. Comparing Fig.8 with Fig.9 one could say that
agreement between $P(r)$ and is better if $\lambda<0,1$ .
\end{center}
\end{figure}

\end{document}